\documentclass[preprint,amsmath,amssymb,amssymb,aps,prl]{revtex4}
\usepackage[colorinlistoftodos, color=green!40, prependcaption]{todonotes}
\usepackage{graphicx,url}
\usepackage[utf8]{inputenc}  
\usepackage[english]{babel}
\usepackage{physics}
\usepackage{float}
\usepackage{url}
\usepackage{soul}
\usepackage{graphicx}
\usepackage{dcolumn}
\usepackage{bm}
\usepackage{times}
\usepackage{mathrsfs}
\usepackage{color}
\usepackage{todonotes}
\usepackage[pdftex, pdftitle={Article}, pdfauthor={Author}]{hyperref}
\usepackage{natbib}
\begin{document}
\title{Cryogenic Dielectric Metasurface-Integrated Superconducting Nanowire Single-Photon Detectors}
\author{Amir Targholizadeh$^{1}$, Grigoriy Y. Nikulin$^{1}$ and Pankaj K. Jha$^{1,2,}$$\footnote{Correspondence and requests for materials should be requested to P.K.J (pkjha@syr.edu)}$}
\affiliation{$^{1}$Quantum Technology Laboratory $\langle \text{Q}|\text{T}|\text{L}\rangle$, Department of Electrical Engineering and Computer Science, Syracuse University, Syracuse, NY 13244, USA}
\affiliation{$^{2}$Institute for Quantum \& Information Sciences, Syracuse University, Syracuse, NY 13244 USA}
\begin{abstract}
Over the past decade, multi-element superconducting nanowire single-photon detectors (SNSPDs) have emerged as the leading single-photon detection technology due to their exceptional system detection efficiency (SDE), ultrahigh timing precision, negligible dark counts, etc. However, achieving these performances with a \textit{single-element} SNSPD has been an outstanding challenge due to a fundamental trade-off: a large active area is necessary for high SDE, while a smaller area is crucial for higher photon count rates, lower dark counts, and lower jitter. Here, we introduce an all-dielectric cryogenic metalens integrated with a single-element SNSPD to achieve high SDE with a smaller active area. Furthermore, we leverage a bifunctional metalens to demonstrate polarization-resolved photodetection at the telecommunication wavelength theoretically. Integrating multifunctional cryogenic metaurfaces with the state-of-the-art SNSPDs may enable novel capabilities with reduced size, weight, power, cost, and cooling (SWaP-C$^{2}$) requirements.
\end{abstract}
\date{\today}
\maketitle

In the last decade, superconducting nanowire single-photon detectors (SNSPDs) have emerged as a leading single-photon detection technology owing to near unity system detection efficiency (SDE)~\cite{Reddy2020,Chang2021}, $\sim$ GHz photon count rate (PCR) ~\cite{Craiciu2023, Resta2023}, dark count rate (DCR) below 10$^{-5}$ counts/s~\cite{Chiles2022}, and $\sim$ ps timing resolution~\cite{Korzh2020}. This remarkable progress has enabled high-rate (1.3 Gb s$^{-1}$)~\cite{Zhang2018} and long-distance quantum key distribution (404 km~\cite{Yin2016}), long-distance quantum communications ($>$ 1200 km)~\cite{Yin2017}, among others. However, optimizing these characteristic device metrics simultaneously within a \textit{single-element} SNSPD remains an outstanding challenge because of the inherent trade-off between them. A typical single-element SNSPD consists of a thin ($\sim$5-10 nm), narrow ($\sim$100 nm), and long wire (a few hundred micrometers) made of a low-temperature superconductor, such as niobium nitride (NbN), niobium titanium nitride (NbTiN), or tungsten silicide (WSi), coiled into a meander pattern covering an active area large enough to capture light from a coupled single-mode optical fiber as well as accommodate slight fiber misalignment~\cite{AA, Natarajan2011, Yamashita2017, Holzman2019, Zadeh2021, Miller2011}. Achieving high PCR, low DCR, and low timing jitter requires a shorter nanowire. However, as the nanowire length decreases, the active area shrinks, which in turn reduces the SDE critical for  \textquotedblleft photon-starved\textquotedblright\, applications~\cite{Hall2015, Shin2016, Erkmen2010, Farr2013}. 

A straightforward approach to considerably reduce the length of the nanowire without compromising the active area is the implementation of an interleaved design~\cite{Dauler2009}. This method substitutes a single elongated nanowire with an array of multiple shorter nanowires sharing the same active area~\cite{Rosenberg2013, Huang2018, Zhang2019, Wollman2019}. However, as the number of interleaved nanowires increases, within a given active area, cross-talk between the nanowires due to electrical or thermal coupling becomes detrimental. Additionally, because each nanowire operates independently, they need separate biasing, readout circuits, and other components, which can significantly increase the complexity of the readout system and the thermal load of the detector, creating challenges for scalability~\cite{Steinhauer2021, Oripov2023, Gao2025}. Another approach is to efficiently expand the collection area with a conventional lens while maintaining a smaller active area~\cite{Bellei2016}. This approach eliminates the need for a single-mode fiber to couple light to the SNSPD, thereby overcoming the limitation of the minimum active area imposed by the fiber's core diameter in conventional fiber-coupled SNSPDs. Furthermore, free-space coupled SNSPDs are especially promising for long-wave infrared (LWIR) applications where single-mode fibers are more rigid and fragile. Recently, a single-element and lensed NbN SNSPD exhibited an SDE 72\% $\pm$ 3.7 \% at 1550 nm with sub-0.1 Hz DCR, and sub-15 ps timing jitter~\cite{Mueller2021}. However, the full potential of this approach remains unrealized because it relies on conventional optical elements that are large, bulky, and less congruent with integrated quantum photonic systems.

In this Letter, we propose and demonstrate a long-sought-after solution to a miniaturized single-element SNSPD with high SDE while preserving the benefits of a shorter active area. We harness the exceptional optical wavefront molding capabilities of an array of judiciously designed dielectric nanoantennas, also known as a metasurface, to serve as a flat lens that significantly enhances the effective collection area of a single-element SNSPD, thereby enabling high SDE of a larger detector with a smaller active area. Furthermore, we designed a polarization beam-splitting metalens to demonstrate polarization-resolved single-photon detection at 1550 nm using a pair of single-element SNSPDs.

Using dielectric metasurfaces to miniaturize single-element SNSPDs offers several advantages. First, metasurfaces enable the simultaneous control of key light properties—phase, amplitude, and polarization—providing a promising way to replace bulky optical components with flat, ultrathin, and lightweight alternatives that are CMOS-compatible~\cite{Lin2014, Chen2016, Kamali2018, Genevet2017, Froch2025}. Second, metasurfaces offer flat optical components for LWIR wavelengths, where larger-core optical fibers necessitate larger active areas for SNSPDs and thus suffer from low PCR, high DCR, and poor timing jitter~\cite{Taylor2019}. Third, a single metasurface can provide complex functionalities that would otherwise require a combination of several conventional bulky optical components to accomplish~\cite{Rubin2019, Fan2023, Jones2022, Swartz2024, Kamali2016}. This may enable novel capabilities in the state-of-the-art SNSPDs with reduced size, weight, power, cost, and cooling requirements.

\textit{Metalensed SNSPD:} Figure 1 illustrates a schematic of an all-dielectric metalens that focuses an incoming $x$-polarized light onto a meander-shaped single-element SNSPD aligned with the polarization direction. Unlike a conventional lens that works by refracting light, a metalens ~\cite{Arbabi2015, Khorasaninejad2016, Wang2018} focuses light by shaping the incident wavefront through abrupt phase changes locally imparted by the nanoantennas. The metalens collects light over an area $A_M$. Then it focuses it onto the SNSPD's active area $A_D$ with efficiency $\eta_{c}$, effectively increasing the light-collection area from $A_{D}$ to $\eta_{c}A_{M}$ without requiring a larger active area. The SNSPD is composed of a 8 nm thick and 100 nm wide niobium-titanium nitride (NbTiN) meander nanowire with a 50 percent fill fraction and covers an active area $A_{D}$ =  5$\times$5 $\mu$m$^{2}$, which is about an order of magnitude smaller than a typical SNSPD operating at 1550 nm wavelength ~\cite{Reddy2020}. Without loss of generality, we considered NbTiN as the superconducting material for photodetection, as various groups have experimentally demonstrated high efficiency, low timing jitter, low dark-count rate, and high photon-detection rates with NbTiN-based SNSPDs ~\cite{Yang2019,Azem2024}. The NbTiN nanowires are integrated with a simple cavity structure, consisting of a SiO$_2$ spacer layer and a gold back-reflector to optimize photoabsorption~\cite{Rosfjord2006, Anant2008}. The absorptance $A$ of the nanowire is given by~\cite{Anant2008}.
\begin{equation}
\label{eq:absorptance}
A = \frac{\displaystyle\int_V \omega\,\Im[\varepsilon(\mathbf{r})]\,|\mathbf{E}(\mathbf{r})|^2\mathrm{d}V}
{\displaystyle\frac{1}{\eta_0}\int_S|\mathbf{E}_{\mathrm{inc}}|^2\,\mathrm{d}S},
\end{equation}
where $\omega$ is the optical angular frequency, $\Im[\varepsilon(\mathbf{r})]$ is the imaginary component of the NbTiN permittivity~\cite{baner2018}, $\mathbf{|E}(\mathbf{r})|^2$ is the time-averaged local electrical intensity within the nanowires, $\mathbf{E}_{\mathrm{inc}}$ is the amplitude of the incident field, and $\eta_0$ is the impedance of free space. Our simulation shows that without the cavity, the absorptance of an infinite NbTiN grating on a SiO$_2$ substrate is 23\% for a plane-wave at normal incidence, and the absorptance increases to $>$ 92\% with a cavity with a 233 nm thick SiO$_2$ layer (See Supporting Information section 1 Fig. S1(a)). 

The phase profile across the metalens is given by~\cite{Arbabi2015, Khorasaninejad2016, Wang2018} 
\begin{equation}
\varphi(x,y) = -\frac{2\pi}{\lambda}\left(\sqrt{x^{2}+y^{2}+f^{2}}-f\right)
\end{equation}
where $\lambda$ is the wavelength in vacuum, and  $f$ is the focal length of the metalens. For the designed wavelength of 1550 nm, our metalens is composed of an array of amorphous silicon ($a$-Si) cylindrical nanopillars (as shown in Fig. 2 (a)) of height $H$ = 850 nm on a silicon dioxide (SiO$_{2}$) substrate with a square lattice of pitch $P$ = 600 nm. The refractive index of $a$-Si at 5 K was taken from~\cite{komma2012}. By changing the diameter of the nanopillars, the phase of the transmitted light is molded to span the entire $2\pi$ phase range. We mimicked the metalens profile with a set of eight different nanocylinder phase shifters, yielding a transmission efficiency of $90\%$ of the metalens (See Supporting Information section 2 Fig. S2(a)). To quantify the performance of our metalensed SNSPD, we used the quantitative description of the system detection efficiency (SDE) defined as $\eta_{\text{SDE}} = \eta_{\text{c}} \times \eta_{\text{a}} \times \eta_{\text{r}}$ where $\eta_{\text{c}}, \eta_{\text{a}}$, and $\eta_{\text{r}}$ are the coupling efficiency, absorption efficiency, and registering efficiency, respectively ~\cite{Natarajan2011}. Assuming that $\eta_{\text{r}}$ remains the same for a metalensed and reference (without metalens, free-spaced) detector, and $\eta_{\text{c}}$ of the reference detector is unity, the enhancement factor in the SDE of a metalensed SNSPD compared to the reference detector can be defined as~\cite{Xu2021} 
\begin{equation}
\frac{\eta_{\text{SDE, M}}}{\eta_{\text{SDE, R}}} = \left(\frac{\tilde{\eta}_{\text{c}}A_{M}}{A_{D}}\right)\frac{\tilde{\eta}_{a}}{\eta_{a}}
\end{equation}

Choosing an optimal numerical aperture (NA) of the metalens is crucial for a small detector area $A_{D}$ = 5$\times$5$\,\mu$m$^{2}$. Intuitively, a moderate to high NA metalens would be necessary to focus the incoming light to a spot size (diameter) of less than 5$\,\mu$m. However, as the NA of a metalens increases, it exhibits a gradual decrease in focusing efficiency, $\tilde{\eta}_{c}$ ~\cite{Prather2001,Johansen2024}. Furthermore, the components of the Poynting vector ($S_{x}, S_{y}, S_{z}$) redistribute in the focal plane, affecting the absorption efficiency  $\tilde{\eta}_{\text{a}}$. Figure 2(b) shows the plot of the enhancement factor $\eta_{\text{SDE, M}}/\eta_{\text{SDE, R}}$ as a function of the NA of the metalens. For these simulations, we fixed the area of the metalens $A_{M}$ = 40 $\times$ 40 $\mu$m$^{2}$ (to keep the metalens footprint small and computationally light) and varied the focal length \textit{f} = 97.98 $\mu$m to 9.69 $\mu$m. This range of focal length spanned the NA from 0.2 to 0.9. It is well-known that metalens exhibits NA-dependent focal length shift, i.e., the simulated focal length ($f_{s}$) is different from the designed focal length ($f$) used in Eq. (2)~\cite{Yuxuan2018}. Subsequently, for each NA, the detector is considered at the simulated focal length rather than the designed focal length (See Supporting Information, Section 3, Fig. S3(b)). Our simulations show that a modest-sized metalens can provide an enhancement factor reaching $\sim$ 48  for NA = 0.45, which corresponds to a simulated focal length of $\sim$35.5 $\mu$m. Figure 2(c) shows the focusing effect of the metalens as shown by the field intensity distribution in the $x$-$z$ plane. The focal spot on the NbTiN nanowire ($x$-$y$ plane) is shown in the inset, and the full-width at half-maximum (FWHM) of the intensity profile along the $x$-axis on the focal plane is 1.62 $\mu $m (See Supporting Information section 2 Fig. S2(b)).  

\textit{Multifunctional Metalensed SNSPD:} Polarization is one of the fundamental properties of light, and its detection has diverse applications ranging from biomedical imaging and remote sensing to quantum key distribution and astronomy~\cite{MOS}. Conventional meander-shaped SNSPDs are inherently polarization-sensitive. Our simulations show that NbTiN nanowires, under plane-wave illumination, show polarization dependence: absorptance exceeds $92\%$ for $\mathbf{E}\!\parallel$ wire but drops to $45\%$ for $\mathbf{E}\!\perp$ wire (See Supporting Information section 1 Fig. S1(b)). For circular polarization, the absorptance is $\sim 68 \%$. The degree of polarization $C = \left(N_{\parallel} - N_{\perp}\right)/\left(N_{\parallel} + N_{\perp}\right) = 0.34 $ is low and thereby requires a polarization filter or a polarizer before detection. A unique feature of metasurfaces is the ability to encode more than one function in a single layer~\cite{Chen2020,Liu2021, Kossowski2025}. Here, we designed a dielectric linear polarization beam splitting (LPBS) metalens integrated with a pair of NbTiN SNSPDs oriented mutually orthogonal to demonstrate polarization-resolved single-photon detection at telecommunication wavelength theoretically.  

Figure~3(a-c) shows the schematic illustration of a dielectric LPS metalens, which combines a metalens with a linear polarization beam splitter. Here, incident circularly polarized light is decomposed into its $x$- and $y$-components, which are then focused onto SNSPDs aligned with the  $x$- and $y$-axes, respectively. Similarly, if the LPS metalens is illuminated with $x$-polarized ($y$-polarized) light, the light gets focused on the $x$-aligned ($y$-aligned) SNSPD. Taking the metalens center as the origin $(0,0,0)$, Eq.~(2) gives the phase profile for an on-axis focus at $(0,0,f)$. In our design, this phase is modified to produce two laterally displaced, cofocal spots at $(-x_{0},0,f)$ and $(x_{0},0,f)$, routing the $x$-polarized component to the negative-$x$ focus and the $y$-polarized component to the positive-$x$ focus. The corresponding aperture-plane phases are $\varphi_{x,y}(x,y)=-(2\pi/\lambda)\left(\sqrt{(x\pm x_{0})^{2}+y^{2}+f^{2}}-f\right)$ whose equiphase contours are circular and centered at $(-x_0,0)$ for $\varphi_x$ and at $(x_0,0)$ for $\varphi_y$; see Figs.~3(b,c). Both phase maps are realized on a single metasurface by exploiting form birefringence: rectangular $a$-Si nanobricks on a low-index SiO$_2$ substrate offer different effective indices for the TE and TM modes, allowing for independent phase control of orthogonal polarizations. At $\lambda=1550\,\mathrm{nm}$, sweeping the in-plane widths $(w_x,w_y)$ yields full $0$-$2\pi$ phase coverage for both polarizations with high transmission. We quantize each polarization’s phase to eight levels over $0$-$2\pi$ in $\pi/4$ increments; the resulting $8\times8$ phase grid requires $64$ phase-pair states. We set the unit-cell pitch to $500$ nm and the nanobrick height to $h=1960$ nm, which is larger than in typical $1550$ nm metalenses~\cite{Hada2024}, to increase modal separation and provide two independent phase degrees of freedom per unit cell. We precompute a library of $64$ nanobrick geometries that map $(\varphi_x,\varphi_y)$ to $(w_x,w_y)$ with mean transmittance $>\!90\%$ (See Supporting Information section 4 Table. S1 and Fig. S4(a)).

Polarization-resolved photodetection can be quantified by crosstalk (XT) between the two ports of photodetection as~\cite{xiong2014}:
\begin{equation}
\label{eq:XTdB}
\mathrm{XT}_{c}\,[\mathrm{dB}] \equiv 10\log_{10}\!\left(\frac{P_{c\to \bar c}}{P_{c\to c}}\right), \qquad c\in\{x,y\},
\end{equation}
where \(\bar c\) denotes the orthogonal partner of \(c\). Here \(P_{c\to c}\) and \(P_{c\to \bar c}\) are the (dimensionless) absorptances of the intended port \(c\) and the orthogonal port \(\bar c\), respectively—the corresponding SNSPDs are designed for those polarizations, with their nanowire meanders oriented along \(c\) or \(\bar c\). These quantities are computed in the FDTD solver (Tidy3D by Flexcompute Inc.)\cite{flex} from Eq.~(1) by integrating the ohmic-loss density over the nanowire volume and normalizing to unit incident power. Figure~4(a) plots \(\mathrm{XT}_{x}\) and \(\mathrm{XT}_{y}\) from Eq.~(\ref{eq:XTdB}) against the center-to-center distance \(d=2x_0\) and identifies an operating point at \(x_0=8~\mu\mathrm{m}\) (\(d=16~\mu\mathrm{m}\)), where \(\mathrm{XT}_{x}< -22~\mathrm{dB}\) and \(\mathrm{XT}_{y}< -16~\mathrm{dB}\). For an ideal metalens the response \(\mathrm{XT}_{x}\) and \(\mathrm{XT}_{y}\) should be identical. However, we observe deviation from this ideal behavior due to the difference in focusing performance of a metalens for $x$ and $y$-polarized light, which is regularly observed in dielectric metalenses (See Supporting Information, Section 4, Fig. S4(b-d)). Figures~4(b–d) show intensity maps of the LPS metalensed SNSPDs for circular, \(x\)-polarized, and \(y\)-polarized inputs at optimum separation distance, respectively: top panels display \(x\)–\(z\) cross sections, and bottom panels show focal-plane \(x\)–\(y\) maps at \(z=f_{s}\).

A viable way to precisely align a metalens at the designated distance from the SNSPD inside an optical cryostat is to mount it on a $z$-axis nanopositioner within the cold shield, along with the SNSPD, which is mounted on an $x$-$y$ nanopositioner. This modular approach provides the flexibility to actively correct for any lateral shift (out-of-focus) in the position of either the metalens or the SNSPD when the cryostat is cooled to 4K. Such corrections are routinely done in an optical cryostat with cryogenic objective lenses~\cite{Jha2021, Akbari2022}. Another approach is to fabricate the metalens directly on top of the SNSPDs, similar to microlenses~\cite{Xu2021}, by incorporating a spacer layer. Here, the alignment can be achieved within the precision limits of the lithography tools. This integrated device would exhibit little to no lateral misalignment when cooled to 4K, thereby maintaining its performance (See Supporting Information, Section 2, Fig. S2(a)). However, it would involve complex nanofabrication, which could reduce the device yield.

\begin{figure}[t]
\centerline{\includegraphics[scale =0.95]{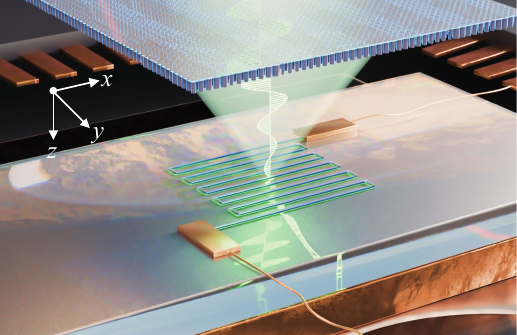}}
  \caption{\textbf{Metalensed single-element SNSPD}. Schematic illustration of an all-dielectric $a$-Si metalens that effectively enhances the collection area of a miniaturized single-element SNSPD by capturing incoming light over the metalens area $A_{M}$ and focusing it onto the active area $A_{D}$. The meander-shaped SNSPD is integrated with an optical cavity to optimize the absorptance of the focused light. With a modest size (area) of a metalens, the system detection efficiency of the metalensed SNSPD is 48-fold higher compared to a reference SNSPD (without a metalens) for a given detector size.}
 \end{figure}
 
 \clearpage
 \begin{figure}[t]
\centerline{\includegraphics[scale =0.9]{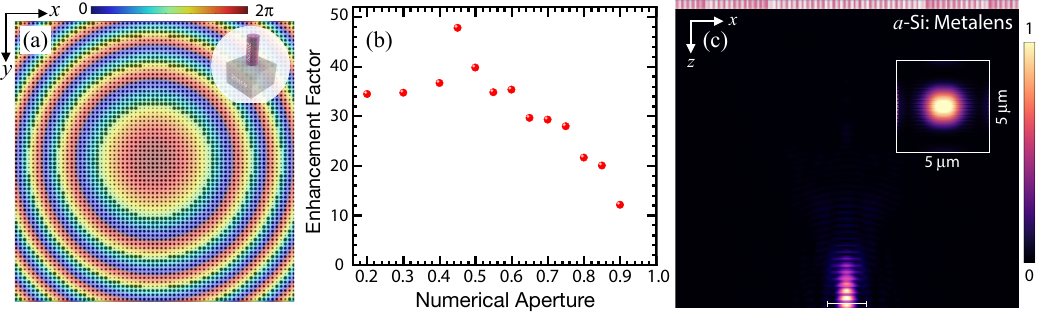}}
  \caption{\textbf{Metalensed NbTiN SNSPD SDE enhancement.}\,(a) For the designed wavelength of 1550 nm, our metalens used a set of eight nanopillars with radius of 0.180 $\mu$m, 0.193 $\mu$m, 0.214 $\mu$m, 0.242 $\mu$m, 0.102 $\mu$m, 0.140 $\mu$m, 0.157 $\mu$m, 0.168 $\mu$m) corresponding to $\pi/8, 3\pi/8, 5\pi/8, 7\pi/8, 9\pi/8, 11\pi/8, 13\pi/8, 15\pi/8$ phase shifts, respectively. The height of each nanopillar is set to 850 nm, and the pitch is 600 nm. (b) Plot of the enhancement factor of the system detection efficiency of the metalensed SNSPD defined as $\eta_{\text{SDE, M}}/\eta_{\text{SDE, R}}$ against the numerical aperture of the metalens of size $A_{M}$ = 40 $\times$40 $\mu$m$^{2}$. (c) Plot of intensity distribution of the metalens of NA 0.45 and illuminated by an $x$-polarized light in the $x$-$z$ plane. The inset shows the intensity distribution on the NbTiN nanowires $x$-$y$ plane.}
 \end{figure}

\clearpage
\begin{figure}[t]
\centerline{\includegraphics[scale =0.9]{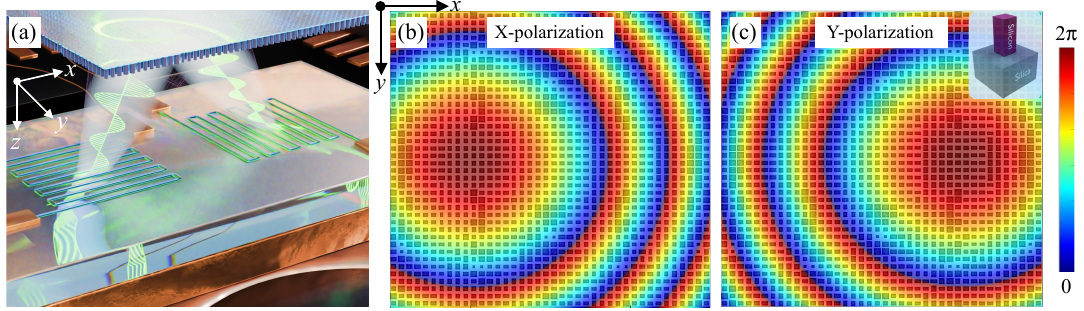}}
  \caption{\textbf{Bifunctional metalensed single-element SNSPDs.} (a) Schematic illustration of a dielectric LPBS metalensed SNSPDs. An incident circularly polarized beam is decomposed into its $x$- and $y$-polarized components, which are focused onto spatially separated meandered SNSPDs oriented along the $x$- and $y$-axes, respectively. For linearly polarized input, an $x$- ($y$-) polarized beam is focused onto the $x$- ($y$-) oriented SNSPD. (b–c) The LPBS metalens design employs superimposed phase profiles corresponding to off-axis focusing for the two orthogonal polarizations.}
 \end{figure}
 
\clearpage
\begin{figure}[t]
\centerline{\includegraphics[scale =0.9]{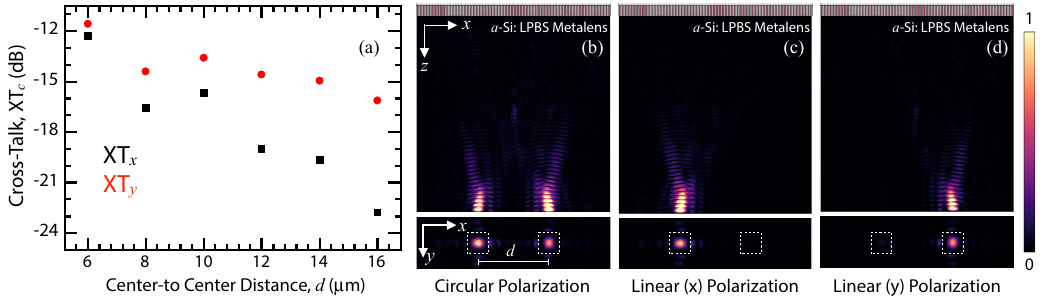}}
  \caption{\textbf{Polarization-resolved photodetection using LPBS metalensed SNSPDs.} (a) Crosstalk (XT$_{c}$) for the \(x\)- and \(y\)-ports versus center-to-center distance \(d=2x_0\) between the detectors as defined in Eq.~(\ref{eq:XTdB}).(b–d) Field–intensity maps of the LPBS metalensed SNSPD at \(d=16~\mu\mathrm{m}\) for circular, \(x\)–linear, and \(y\)–linear inputs, respectively. \emph{Top:} meridional \(x\)–\(z\) slices showing off–axis focusing to \(\pm x_0\). \emph{Bottom:} focal–plane \(x\)–\(y\) maps at \(z=f_{s}\).}
 \end{figure}


\clearpage
\begin{thebibliography}{99}
\bibitem{Reddy2020} D. V. Reddy, R. R. Nerem, S. W. Nam, R. P. Mirin, and V. B. Verma, Superconducting nanowire single-photon detectors with 98\% system detection efficiency at 1550 nm, Optica \textbf{7}, 1649 (2020).
\bibitem{Chang2021} J. Chang, J. W. N. Los, J. O. Tenorio-Pearl, N. Noordzij, R. Gourgues, A. Guardiani, J. R. Zichi, S. F. Pereira, H. P. Urbach, V. Zwiller, S. N. Dorenbos, and I. Esmaeil Zadeh, Detecting telecom single photons with (99.5$^{+0.5}_{-2.07}$)\% system detection efficiency and high time resolution, APL Photonics \textbf{6}, 036114 (2021).
\bibitem{Craiciu2023} I. Craiciu, B. Korzh, A. D. Beyer, A. Mueller, J. P. Allmaras, L. Narvaez, M. Spiropulu, B. Bumble, T. Lehner, E. E. Wollman, and M. D. Shaw, High-speed detection of 1550 nm single photons with superconducting nanowire detectors, Optica \textbf{10}, 183 (2023).
\bibitem{Resta2023} G. V. Resta, L. Stasi, M. Perrenoud, S. El-Khoury, T. Brydges, R. Thew, H. Zbinden, and F. Bussières, Gigahertz detection rates and dynamic photon-number resolution with superconducting nanowire arrays, Nano Lett. \textbf{23}, 6018 (2023).
\bibitem{Chiles2022} J. Chiles, I. Charaev, R. Lasenby, M. Baryakhtar, J. Huang, A. Roshko, G. Burton, M. Colangelo, K. V. Tilburg, A. Arvanitaki, S. W. Nam, and K. K. Berggren, New constraints on dark photon dark matter with superconducting nanowire detectors in an optical haloscope, Phys. Rev. Lett. \textbf{128}, 231802 (2022).
\bibitem{Korzh2020} B. Korzh, Q. Y. Zhao, J. P. Allmaras, S. Frasca, T. M. Autry, E. A. Bersin, A. D. Beyer, R. M. Briggs, B. Bumble, M. Colangelo, G. M. Crouch, A. E. Dane, T. Gerrits, A. E. Lita, F. Marsili, G. Moody, C. Pena, E. Ramireza, J. D. Rezac, N. Sinclair, M. J. Stevens, A. E. Velasco, V. B. Verma, E. E. Wollman, S. Xie, D. Zhu, P. D. Hale, M. Spiropulu, K. L. Silverman, R. P. Mirin, S. W. Nam, A. G. Kozorezov, M. D. Shaw, and K. K. Berggren, Demonstration of sub-3 ps temporal resolution with a superconducting nanowire single-photon detector, Nat. Photonics \textbf{14}, 250 (2020).
\bibitem{Zhang2018} Z. Zhang, C. Chen, Q. Zhuang, F. N. C. Wong, and J. H. Shapiro, Experimental quantum key distribution at 1.3 gigabit-per-second secret-key rate over a 10 dB loss channel, Quantum Sci. Technol. \textbf{3}, 025007 (2018).
\bibitem{Yin2016} H. L. Yin, T. Y. Chen, Z. W. Yu, H. Liu, L. X. You, Y. H. Zhou, S. J. Chen, Y. Mao, M. Q. Huang, W. J. Zhang, H. Chen, M. J. Li, D. Nolan, F. Zhou, X. Jiang, Z. Wang, Q. Zhang, X. B. Wang, and J. W. Pan, Measurement-device-independent quantum key distribution over a 404 km optical fiber, Phys. Rev. Lett. \textbf{117}, 190501 (2017).
\bibitem{Yin2017} J. Yin, Y. Cao, Y. H. Li, S. K. Liao, L. Zhang, J. G. Ren, W. Q. Cai, W. Y. Liu, B. Li, H. Dai, G. B. Li, Q. M. Lu, Y. H. Gong, Y. Xu, S. L. Li, F. Z. Li, Y. Y. Yin, Z. Q. Jiang, M. Li, J. J. Jia, G. Ren, D. He, Y. L. Zhou, X. X. Zhang, N. Wang, X. Chang, Z. C. Zhu, N. Le Liu, Y. A. Chen, C. Y. Lu, R. Shu, C. Z. Peng, J. Y. Wang, and J. W. Pan, Satellite-based entanglement distribution over 1200 kilometers, Science \textbf{356}, 1140 (2017).
\bibitem{AA} The active area is defined as the region capturing both nanowires and the space between them.
\bibitem{Natarajan2011} C. M. Natarajan, M. G. Tanner, and R. H. Hadfield, Superconducting nanowire single-photon detectors: Physics and applications, Supercond. Sci. Technol. \textbf{25}, 063001 (2011).
\bibitem{Yamashita2017} T. Yamashita, S. Miki, and H. Terai, Recent progress and application of superconducting nanowire single-photon detectors, IEICE Trans. Electron. \textbf{E100-C}, 274 (2017).
\bibitem{Holzman2019} I. Holzman and Y. Ivry, Superconducting nanowires for single-photon detection: Progress, challenges, and opportunities, Adv. Quantum Technol. \textbf{2}, 1800058 (2019).
\bibitem{Zadeh2021} I. E. Zadeh, J. Chang, J. W. N. Los, S. Gyger, A. W. Elshaari, S. Steinhauer, S. N. Dorenbos, and V. Zwiller, Superconducting nanowire single-photon detectors: A perspective on evolution, state-of-the-art, future developments, and applications, Appl. Phys. Lett. \textbf{118}, 190502 (2021).
\bibitem{Miller2011} A. J. Miller, A. E. Lita, B. Calkins, I. Vayshenker, S. M. Gruber, and S. W. Nam, Compact cryogenic self-aligning fiber-to-detector coupling with losses below one percent, Opt. Express \textbf{19}, 9102 (2011).
\bibitem{Hall2015} D. J. Hall, N. Bush, N. Murray, J. Gow, A. Clarke, R. Burgon, A. Holland, Challenges in photon-starved space astronomy in a harsh radiation environment using CCDs, Proc. SPIE 9602, 96020U (2015).
\bibitem{Shin2016} D. Shin, F. Xu, D. Venkatraman, R. Lussana, F. Villa, F. Zappa, V. K. Goyal, F. N. C. Wong, J. H. Shapiro, Photon-efficient imaging with a single-photon camera, Nat. Commun. \textbf{7}, 12046 (2016).
\bibitem{Erkmen2010} B. Erkmen, B. Moision, and K. Birnbaum, A review of the information capacity of single-mode free-space optical communication, Proc. SPIE \textbf{7587}, 75870N (2010).
\bibitem{Farr2013} W. H. Farr, J. M. Choi, and B. Moision, 13 bits per incident photon optical communications demonstration, Proc. SPIE \textbf{8610}, 861006 (2013).
\bibitem{Dauler2009} E. A. Dauler, A. J. Kerman, B. S. Robinson, J. K. W. Yang, B. Voronov, G. Goltsman, S. A. Hamilton, and K. K. Berggren, Photon-number-resolution with sub-30-ps timing using multi-element superconducting nanowire single photon detectors, J. Mod. Opt. \textbf{56}, 364 (2009).
\bibitem{Rosenberg2013} D. Rosenberg, A. J. Kerman, R. J. Molnar, and E. A. Dauler, High-speed and high-efficiency superconducting nanowire single photon detector array, Opt. Express \textbf{21}, 1440 (2013).
\bibitem{Huang2018} J. Huang, W. Zhang, L. You, C. Zhang, C. Lv, Y. Wang, X. Liu, H. Li, and Z. Wang, High-speed superconducting nanowire single-photon detector with nine interleaved nanowires, Supercond. Sci. Technol. \textbf{31}, 074001 (2018).
\bibitem{Zhang2019} W. Zhang, J. Huang, C. Zhang, L. You, C. Lv, L. Zhang, H. Li, Z. Wang, and X. Xie, A 16-pixel interleaved superconducting nanowire single-photon detector array with a maximum count rate exceeding 1.5 GHz, IEEE Trans. Appl. Supercond. \textbf{29}, 2200204 (2019).
\bibitem{Wollman2019} E. E. Wollman, V. B. Verma, A. E. Lita, W. H. Farr, M. D. Shaw, R. P. Mirin, and S. W. Nam, Kilopixel array of superconducting nanowire single-photon detectors, Opt. Express \textbf{27}, 35279 (2019).
\bibitem{Steinhauer2021} S. Steinhauer, S. Gyger, and V. Zwiller, Progress on large-scale superconducting nanowire single-photon detectors, Appl. Phys. Lett. \textbf{118}, (2021).
\bibitem{Oripov2023} B. G. Oripov, D. S. Rampini, J. Allmaras, M. D. Shaw, S. W. Nam, B. Korzh, and A. N. McCaughan, A superconducting nanowire single-photon camera with 400{,}000 pixels, Nature \textbf{62}, 730 (2023).
\bibitem{Gao2025} J. Gao, J. Chang, B. L. Rodriguez, I. E. Zadeh, V. Zwiller, and A. W. Elshaari, From pixels to camera: Scaling superconducting nanowire single-photon detectors for imaging at the quantum-limit, arXiv:2505.24725 (2025).
\bibitem{Bellei2016} F. Bellei, A. P. Cartwright, A. N. McCaughan, A. E. Dane, F. Najafi, Q. Zhao, and K. K. Berggren, Free-space-coupled superconducting nanowire single-photon detectors for infrared optical communications, Opt. Express \textbf{24}, 3248 (2016).
\bibitem{Mueller2021} A. S. Mueller, Free-space coupled superconducting nanowire single-photon detector with low dark counts, Optica \textbf{8}, 1586 (2021).
\bibitem{Lin2014} D. Lin, P. Fan, E. Hasman, and M. L. Brongersma, Dielectric gradient metasurface optical elements, Science \textbf{345}, 298 (2014).
\bibitem{Chen2016} H.-T. Chen, A. J. Taylor, and N. Yu, A review of metasurfaces: Physics and applications, Rep. Prog. Phys. \textbf{79}, 076401 (2016).
\bibitem{Kamali2018} S. M. Kamali, E. Arbabi, A. Arbabi, and A. Faraon, A review of dielectric optical metasurfaces for wavefront control, Nanophotonics \textbf{7}, 1041 (2018).
\bibitem{Genevet2017} P. Genevet, F. Capasso, F. Aieta, M. Khorasaninejad, and R. Devlin, Recent advances in planar optics: From plasmonic to dielectric metasurfaces, Optica \textbf{4}, 139 (2017).
\bibitem{Froch2025} J. E. Fröch, S. Colburn, D. J. Brady, F. Heide, A. Veeraraghavan, and A. Majumdar, Computational imaging with meta-optics, Optica \textbf{12}, 774 (2025).
\bibitem{Taylor2019} G. G. Taylor, D. Morozov, N. R. Gemmell, K. Erotokritou, S. Miki, H. Terai, and R. H. Hadfield, Photon counting LIDAR at 2.3\,$\mu$m wavelength with superconducting nanowires, Opt. Express \textbf{27}, 38147 (2019).
\bibitem{Rubin2019} N. A. Rubin, G. D’Aversa, P. Chevalier, Z. Shi, W. T. Chen, and F. Capasso, Matrix Fourier optics enables a compact full-Stokes polarization camera, Science \textbf{365}, eaax1839 (2019).
\bibitem{Fan2023} Q. Fan, W. Xu, X. Hu, W. Zhu, T. Yue, F. Yan, P. Lin, L. Chen, J. Song, H. J. Lezec, A. Agarwal, Y. Lu, and T. Xu, Disordered metasurface enabled single-shot full-Stokes polarization imaging leveraging weak dichroism, Nat. Commun. \textbf{14}, 7180 (2023).
\bibitem{Jones2022} A. W. Jones, M. Wang, X. Zhang, S. J. Cooper, S. Chen, C. M. Mow-Lowry, and A. Freise, Metasurface-enhanced spatial mode decomposition, Phys. Rev. A \textbf{105}, 053523 (2022).
\bibitem{Swartz2024} B. T. Swartz, H. Zheng, G. T. Forcherio, and J. Valentine, Broadband and large-aperture metasurface edge encoders for incoherent infrared radiation (2024).
\bibitem{Kamali2016} S. M. Kamali, A. Arbabi, E. Arbabi, Y. Horie, and A. Faraon, Decoupling optical function and geometrical form using conformal flexible dielectric metasurfaces, Nat. Commun. \textbf{7}, 11618 (2016).
\bibitem{Arbabi2015} A. Arbabi, Y. Horie, M. Bagheri, and A. F. Faraon, Dielectric metasurfaces for complete control of phase and polarization with subwavelength spatial resolution and high transmission, Nat. Nanotechnol. \textbf{10}, 937 (2015).
\bibitem{Khorasaninejad2016} M. Khorasaninejad, W. T. Chen, R. C. Devlin, J. Oh, A. Y. Zhu, and F. Capasso, Metalenses at visible wavelengths: Diffraction-limited focusing and subwavelength resolution imaging, Science \textbf{352}, 1190 (2016).
\bibitem{Wang2018} S. Wang, P. C. Wu, V-C Su, Y-C Lai, M-K. Chen, H. Y. Kuo, B. H. Chen, Y. H. Chen, T-Z. Huang, J-H Wang, R-M. Lin, C-H. Kuan, T. Li, Z. Wang, S. Zhu, and D. P. Tsai, A broadband achromatic metalens in the visible, Nat. Nanotechnol. \textbf{13}, 227 (2018).
\bibitem{Yang2019} X. Yang, L. You, L. Zhang, C. Lv, H. Li, X. Liu, H. Zhou, and G. Wang, Optimizing the stoichiometry of ultrathin NbTiN films for high-performance superconducting nanowire single-photon detectors, Opt. Express \textbf{27}, 26579 (2019).
\bibitem{Azem2024} A. Azem, D. V. Morozov, D. Kuznesof, C. Bruscino, R. H. Hadfield, L. Chrostowski, and J. F. Young, Mid-infrared characterization of NbTiN superconducting nanowire single-photon detectors on silicon-on-insulator, Appl. Phys. Lett. \textbf{125}, 21 (2024).
\bibitem{Rosfjord2006} K. M. Rosfjord, J. K. W. Yang, E. A. Dauler, A. J. Kerman, V. Anant, B. M. Voronov, G. N. Gol’tsman, and K. K. Berggren, Nanowire single-photon detector with an integrated optical cavity and anti-reflection coating, Opt. Express \textbf{14}, 527 (2006).
\bibitem{Anant2008} V. Anant, A. J. Kerman, E. A. Dauler, J. K. W. Yang, K. M. Rosfjord, and K. K. Berggren, Optical properties of superconducting nanowire single-photon detectors, Opt. Express \textbf{16}, 10750 (2008).
\bibitem{baner2018} A. Banerjee, R. M. Heath, D. Morozov, D. Hemakumara, U. Nasti, I. Thayne, R. H. Hadfield, Optical properties of refractory metal based thin films, Opt. Mater. Express \textbf{8.8}: 2072-2088(2018).
\bibitem{komma2012} J. Komma, C. Schwarz, G. Hofmann, D. Heinert, R. Nawrodt, Thermo-optic coefficient of silicon at 1550 nm and cryogenic temperatures. Appl. Phys. Lett. 101.4 (2012).
\bibitem{Xu2021}Y. Xu, A. Kuzmin, E. Knehr, M. Blaicher, K. Ilin, P-I. Dietrich, W. Freudge, M. Siegel, and C. Koos, Superconducting nanowire single-photon detector with 3D-printed free-form microlenses, Opt. Express \textbf{29}, 27708 (2021).
\bibitem{Prather2001} D. W. Prather, D. Pustai, and S. Shi, Performance of multilevel diffractive lenses as a function of f-number, Appl. Opt. \textbf{40}, 207 (2001).
\bibitem{Johansen2024} V. E. Johansen, U. M. Gür, J. Martínez-Llinás, J. F. Hansen, A. Samadi, M. S. V. Larsen, T. Nielsen, F. Mattinson, M. Schmidlin, N. A. Mortensen, and U. J. Quaade, Nanoscale precision brings experimental metalens efficiencies on par with theoretical promises, Commun. Phys. \textbf{7}, 123 (2024).
\bibitem{Yuxuan2018} J. Yuxuan, Focal shift in metasurface based lenses, Opt. Express \textbf{26}, 8001 (2018).
\bibitem{MOS}M. O. Scully and M. S. Zubairy, \textit{Quantum Optics} (Cambridge University Press, 1997)
\bibitem{Chen2020} S. Chen, W. Liu, Z. Li, H. Cheng, and J. Tian, Metasurface-empowered optical multiplexing and multifunction, Adv. Mater. \textbf{32}, 1805912 (2020).
\bibitem{Liu2021} M. Liu, W. Zhu, P. Huo, L. Feng, M. Song, C. Zhang, L. Chen, H. J. Lezec, Y. Lu, A. Agrawal, and T. Xu, Multifunctional metasurfaces enabled by simultaneous and independent control of phase and amplitude for orthogonal polarization states, Light Sci. Appl. \textbf{10}, 107 (2021).
\bibitem{Kossowski2025} N. Kossowski, Y. Tahmi, A. Loucif, M. Lepers, B. Wattellier, G. Vienne, S. Khadir, and P. Genevet, Metrology of metasurfaces: optical properties, npj Nanophotonics \textbf{2}, 5 (2025).
\bibitem{Hada2024} M. Hada, H. Adegawa, K. Aoki, S. Ikezawa, and K. Iwami, Polarization-separating Alvarez metalens, Opt. Express \textbf{32}, 6672 (2024).
\bibitem{xiong2014} Y. Xiong, D. X. Xu, J. H. Schmid, P. Cheben, S. Janz, W. N. Ye, Fabrication tolerant and broadband polarization splitter and rotator based on a taper-etched directional coupler, Opt. Express 22.14: 17458-17465(2014).
\bibitem{flex} Tidy3D simulation project (2025) \url{https://www.flexcompute.com/tidy3d/qSXRncRYj7}.
\bibitem{Jha2021} P. K. Jha, H. Akbari, Y. Kim, S. Biswas, and H. A. Atwater, Nanoscale axial position and orientation measurement of hexagonal boron nitride quantum emitters using a tunable nanophotonic environment, Nanotechnology \textbf{33}, 015001 (2021).
\bibitem{Akbari2022} H. Akbari, S. Biswas, P. K. Jha, J. Wong, B. Vest, H. A. Atwater, Lifetime-Limited and Tunable Quantum Light Emission in h-BN via Electric Field Modulation, Nano Lett. \textbf{22}, 7798 (2022).

\clearpage
\end{thebibliography}
\end{document}